\documentstyle[preprint,aps]{revtex}
\begin{document}
\draft
\bibliographystyle{prsty}

\title{1D generalized statistics gas: A gauge theory approach}
\author{Silvio J. Rabello\thanks{e-mail: rabello@if.ufrj.br}}
\address{\it Instituto de F\'\i sica, Universidade Federal do Rio de
Janeiro, Rio de Janeiro,  RJ \\ Caixa Postal 68.528-CEP 21945-970, Brazil}

\maketitle
\begin{abstract}

{\sl A field theory with generalized statistics in one space dimension
is introduced. The statistics enters the scene through the coupling of the
matter fields to a statistical gauge field, as it happens in the
Chern-Simons theory in two dimensions. We study the particle-hole excitations
and show that the long wave length physics of this model describes  a
gas obeying the Haldane generalized exclusion statistics.
The statistical interaction is found to provide a way to describe
the low-T critical properties of one-dimensional non-Fermi liquids.}
\end{abstract}

In the Landau theory of  Fermi Liquids \cite{Landau} the interaction effects
are treated as perturbations and we are introduced to the concept of
quasiparticles  having a one-to-one correspondence with the single
particle states of the ideal Fermi gas. A departure from this Fermi
liquid picture can be observed in the study of strongly correlated electron
systems, where even for weak interaction  the Fermi surface is drastically
altered. For many years the theoretical ground for these studies has been
the  one-dimensional (1D) electron liquid described by the Luttinger model
\cite{Lutt}\cite{Hald1}. The discovery of the fractional quantum Hall
effect (FQHE) \cite{Tsui} in 2D systems gave further momentum to the study
of non-Fermi liquids, since it cannot be explained in  terms of single
particle states. In fact the edge excitations of a FQHE sample are believed
to be described by a 1D non-Fermi liquid, based on the (chiral) Luttinger
model \cite{Wen}.

It was recently proposed \cite{Wu1} that by bosonization
of an ideal gas  obeying a generalized exclusion statistics
\cite{Hald2}\cite{Wu2} we may describe the low-T fixed points of 1D non-Fermi
liquids. Instead of being based on the monodromy properties of the wave
functions, this generalized exclusion statistics is based on the variation of
the number of available single particle states as the number of particles in
the system varies, through the relation $\Delta d=-\kappa\Delta N$, with $d$
being the number of available states, N the number of particles and $\kappa$
represents the ``statistical interaction'' parameter. For $\kappa=0$ we have
bosons and for $\kappa=1$ we have fermions, for other values we say that
we are dealing with generalized statistics. In this paper we propose another
point of view for this correspondence between generalized statistics and the
Luttinger model. Inspired by  the success of the
Chern-Simons-Ginsburg-Landau description for the  FQHE in 2D \cite{Zhang},
where the (spinless) electrons are described by bosonic fields coupled to a
Chern-Simons gauge field in a way that the effective theory is fermionic, we
recently introduced \cite{Ra} a 1D gauge field theory that when coupled to the
matter fields has the property of transmuting the statistics of the elementary
quanta. Using this gauge model we here explore the long wave length physics
and find that for the particle-hole (density) fluctuations, the
statistical parameter $\kappa$ can be used to relate the density-density
correlation function of this model to the ones found for the 1D Thirring
and Luttinger models.

We now proceed to describe our gauge model. Imagine a gas of bosonic and
spinless nonrelativistic particles, constrained to move on an infinite
line and that in the many-body language are  described as the elementary
quanta of a complex matter field $\Psi( x,t)$. The only interaction present
here is a ``statistical'' gauge interaction. The Lagrangian density for this
system is
\begin{eqnarray}
\label{1da}
{\cal L}=i\Psi^* (\partial_t+i\phi)\Psi +{1\over 2m}\Psi^*(\partial_x -i
\partial_x \xi+{i\over 2} \chi)^2\Psi
-{1\over 2\kappa\pi}(\chi\partial_t \xi+ \phi\chi)\,,
\end{eqnarray}
with  $\kappa$ a real parameter. It is easy to see that this Lagrangian
density is invariant under the gauge
transformations
\begin{eqnarray}
\label{gt1}
\Psi'(x,t)&=&\Psi(x,t)e^{i\Lambda(x,t)}\,,
\qquad  \Psi'^*(x,t)=\Psi^*(x,t)e^{-i\Lambda(x,t)}\\
\label{gt2} \xi'(x,t)&=& \xi(x,t)+\Lambda(x,t)\,,\qquad
\chi'(x,t)=\chi(x,t)\\ \label{gt3}
\phi'(x,t)&=&\phi(x,t)-\partial_t\Lambda(x,t)\,.
\end{eqnarray}
The two scalar gauge fields $\xi$ and $\chi$ enter the $x-$component of the
covariant derivative in a combination that transforms as a vector potential.
The last term in ${\cal L}$ gives the dynamics of the statistical gauge
fields.

To quantize the above model we follow \cite{Boya} where a gauge invariant
treatment of the two-dimensional  Chern-Simons model can be found.
{}From the symplectic structure of (\ref{1da}) we have the following  equal
time canonical commutation relations (nonzero part)
\begin{equation}
\label{ccr}
[{\hat\Psi}(x,t),{\hat\Psi}^\dagger(y,t)]=\delta(x-y)\,,\qquad
[{\hat \xi}(x,t),{\hat \chi}(y,t)]=-{i2\kappa\pi}\delta( x- y)\,.
\end{equation}
The Hamiltonian operator is given by
\begin{equation}
\label{H}
{\hat H}=\int_{-\infty}^\infty dx\biggl\{
-{1\over 2m}{\hat \Psi}^\dagger[\partial_x -i
(\partial_x {\hat\xi} -{1\over 2}{\hat \chi})]^2{\hat\Psi}
+{\hat \phi}({1\over 2\kappa\pi}{\hat \chi}
+{\hat \Psi}^\dagger{\hat\Psi})\biggr\}\,.
\end{equation}
As in electrodynamics the term that multiplies ${\hat \phi}$ in ${\hat H}$
is the generator of time independent gauge transformations and can be
set identically zero if we work with gauge invariant matter fields, i.e.
fields that commute with ${\hat G}={1\over 2\kappa\pi}{\hat \chi}
+{\hat\Psi}^\dagger{\hat\Psi}$. These fields are given by the gauge invariant
operators
\begin{equation}
\label{phys}
{\hat\Phi}(x,t)={\hat\Psi}(x,t)e^{-i{\hat \xi}(x,t)}\,,\qquad
{\hat\Phi}^\dagger (x,t)={\hat\Psi}^\dagger(x,t)e^{i{\hat \xi}(x,t)}\,.
\end{equation}
With these operators and setting ${\hat G}=0$, the Hamiltonian reads
\begin{equation}
\label{ham}
{\hat H}=-\int_{-\infty}^\infty dx\,
{1\over 2m}{\hat\Phi}^\dagger(\partial_x
-i{\kappa\pi}{\hat\Phi}^\dagger{\hat\Phi})^2{\hat\Phi}\,.
\end{equation}
To study the content of the above ${\hat H}$ we introduce the
occupation number operator
\begin{equation}
\label{num}
{\hat N}=\int_{-\infty}^{\infty} dx\, {\hat\Phi}^\dagger{\hat\Phi}
\end{equation}
and construct the following arbitrary eigenstate of ${\hat N}$
\begin{equation}
\label{ket}
\vert N\rangle=\int dx_1\dots dx_N\,\psi(x_1\dots x_N)
{\hat\Phi}^\dagger(x_1)\dots{\hat\Phi}^\dagger(x_N)\vert 0\rangle\,.
\end{equation}
With $\psi$ being an arbitrary function symmetric under exchange of
any of its entries and $\vert 0\rangle$ is the ground state defined
by ${\hat\Phi}(x)\vert 0\rangle=0$.  If we further insist that this
state is an eigenstate of ${\hat H}$ with eigenvalue $E$, we have that
$\psi$ satisfies the N-body Schr\"{o}dinger equation
\begin{equation}
\label{Sch}
-{1\over 2m}\sum_{a=1}^N\biggl[{\partial\over\partial x_a}
-i{\kappa\pi}
\sum_{b\neq a}\delta(x_a-x_b)\biggr]^2\psi(x_1\dots x_N)=
E\psi(x_1\dots x_N)\,.
\end{equation}
This $\delta -$function interaction was considered in \cite{Kleinert} for
the particular case $\kappa=1$ in order to describe fermions in the
Feynman path integral. One can remove this interaction with aid of a
gauge transformation in $\psi$:
\begin{equation}
\label{psi}
{\bar \psi}(x_1\dots x_N)=exp\biggl(-i{\kappa\pi}\sum_{a<b}
\theta_H(x_a-x_b)\biggr)\psi(x_1\dots x_N)\,,
\end{equation}
where $\theta_H$ is the Heaviside step function, so that ${\bar\psi}$
satisfies the free Schr\"{o}dinger equation
\begin{equation}
\label{Sch2}
-{1\over 2m}\sum_{a=1}^N{\partial^2\over\partial x^2_a}{\bar \psi}
(x_1\dots x_N)=E{\bar \psi}(x_1\dots x_N)\,,
\end{equation}
but obeys a nontrivial condition under exchange of any two arguments
\begin{equation}
\label{anyon}
{\bar \psi}(x_1\dots x_a,x_b\dots x_N)=e^{i{\kappa\pi}\,
sign(x_a-x_b)}{\bar \psi}(x_1\dots x_b,x_a\dots x_N)\,.
\end{equation}
As we can see from (\ref{anyon}) our model leads to  an effective
theory where the elementary quanta display  a generalized  statistics.

Within this model we now proceed to study the behavior of the
collective modes corresponding to the density fluctuations
(particle-hole excitations) of a 1D gas with generalized statistics. For
that purpose we  decompose the matter field $\Psi$ in a density and phase
parts
\begin{equation}
\label{psi2}
\Psi=\sqrt{\rho} e^{i\eta}\,,
\end{equation}
and introduce an uniform background density $-{\bar\rho}$, so
that (\ref{1da}) reads
\begin{eqnarray}
\label{L1}
{\cal L}&=&-\rho\partial_t\eta-(\rho-{\bar\rho})\phi-{1\over 2m}
\biggl[(\partial_x\sqrt{\rho})^2
+\rho(\partial_x\eta-\partial_x\xi+{1\over2}\chi)^2\biggr]\nonumber\\
&-&{1\over 2\kappa\pi}\chi(\phi+\partial_t\xi)\,.
\end{eqnarray}
Variation of $\phi$ gives the constraint
$\chi=-2\kappa\pi(\rho-{\bar\rho})$. Introducing the variables
\begin{equation}
\label{phi}
\sigma=\eta-\xi\,,\qquad\qquad\qquad \tau=\eta+\xi\,,
\end{equation}
we have that (\ref{L1}) goes to
\begin{equation}
\label{L2}
{\cal L}=-\rho\partial_t\sigma-{1\over 2m}\biggl\{(\partial_x\sqrt{\rho})^2
+\rho\biggl[\partial_x\sigma-\kappa\pi(\rho-{\bar\rho})\biggr]^2\biggr\}\,.
\end{equation}
Notice that the above Lagrangian depends only on the
gauge invariant fields $\sigma$ and $\rho$. We now make one more
change of variables,
\begin{equation}
\label{rho}
\rho={\bar\rho}+\partial_x\theta\,,
\end{equation}
so that the density fluctuations $\rho-{\bar\rho}$  are now expressed as
$\partial_x\theta$. As one can see from the first term in
(\ref{L2}) $\Pi_\theta=-\partial_x\sigma$ is the canonical momentum of
$\theta$ in these new variables. If we go to Fourier
modes $\theta(q)={1\over\sqrt{2\pi}}\int_{-\infty}^{+\infty}
dx\,\theta(x)exp(iqx)$,  the long wave length physics is governed by
the Hamiltonian:
\begin{equation}
\label{lrh}
H\simeq {{\bar\rho}\over 2m}\int_{-\infty}^{+\infty}dq\,
\biggl(\Pi_\theta(q)\Pi_\theta(-q)+q^2\kappa^2\pi^2\theta(q)
\theta(-q)\biggr)\,.
\end{equation}
Performing the canonical transformation:
\begin{eqnarray}
\label{Ct}
\theta&\rightarrow&{1\over\sqrt{\kappa\pi}}\theta\,,\\
\Pi_\theta&\rightarrow&\sqrt{\kappa\pi}\Pi_\theta\,,
\end{eqnarray}
the Hamiltonian reads
\begin{equation}
\label{lrhq}
H\simeq {v_s\over 2}\int_{-\infty}^{+\infty}dq\,\biggl(\Pi_\theta(q)
\Pi_\theta(-q)+q^2\theta(q)\theta(-q)\biggr)\,,
\end{equation}
with $v_s$ being the sound velocity
\begin{equation}
\label{vsound}
v_s=\kappa{{\bar\rho}\pi\over m}=\kappa v_F\,.
\end{equation}
Here $v_F$ is the Fermi velocity for a gas of 1D spinless electrons.
The same result was found in \cite{Wu1} by bosonization of an ideal gas
of particles obeying the Haldane generalized exclusion statistics.

To see the relation of our model with 1D fermion models
we now proceed to compute the ground-state wave functional $\psi_0$
for (\ref{lrhq}), we choose the wave functional to depend on
$\theta$ so that $\Pi_\theta$ acts on it as the functional
derivative $-i{\delta\over\delta\theta}$. It can be shown \cite{Frad} that
 $\vert\psi_0\vert^2$  is equivalent to the 2N-point density-density
correlation function provided we set $\partial_x\theta(x)$ to represent
the density for N particles at $x_a$ and N holes
at $y_a$ ($a=1,2,...N$), i.e.
\begin{eqnarray}
\label{dens}
\theta(x)&=&\sqrt{\kappa\pi}\sum_{a=1}^N\biggl(\theta_H(x-x_a)
-\theta_H(y-y_a)\biggr)\nonumber\\
&=&\sqrt{\kappa\pi}\sum_{a=1}^N{1\over 2\pi i}
\int_{-\infty}^{+\infty}dq\,{1\over q}\biggl(e^{iq(x-x_a)}
-e^{iq(x-y_a)}\biggr)\,.
\end{eqnarray}
The solution for the ground-state wave functional is
\begin{equation}
\label{psi0}
\psi_0[\theta]={\cal N} e^{{1\over 2}\int_{-\infty}^{+\infty}dq\,\vert
q\vert \theta(q)\theta(-q)}\,,
\end{equation}
with ${\cal N}$ a normalization constant. If we now extract from
(\ref{dens}) $\theta(q)$ and insert it  in (\ref{psi0}) we obtain
\begin{equation}
\label{Psi1}
\vert\psi_0\vert^2={\cal N}{\prod_{a< b} \vert x_a-x_b\vert^{2\kappa}
\vert y_a-y_b\vert^{2\kappa}\over\prod_{a,b}\vert x_a-y_b
\vert^{2\kappa}}\,.
\end{equation}
For $\kappa=1$ we recover the 2N-point density-density correlation function
for free gapless 1D Dirac fermions \cite{Z-J}, and for $\kappa$
generic (but $>0$) we have the equivalence with the gapless
Thirring and Luttinger models \cite{Frad}. In fact, the use of generalized
statistics to solve the Thirring model can be found in \cite{Klaiber}.

To study the non-ideal gas case, one can introduce a two-body interaction in
(\ref{lrh}) through  the term
\begin{eqnarray}
\label{Hint}
H_{int}&=&{1\over 2}\int dx\,\int dy\,(\rho(x)-{\bar\rho})V(x-y)(\rho(y)
-{\bar\rho})\nonumber\\
&=&\sqrt{\pi\over 2}\int_{-\infty}^{+\infty}dq\,V(q)q^2\theta(q)\theta(-q)\,.
\end{eqnarray}
If the two-body potential is sufficiently short ranged, so that in the
low $q$ approximation we keep only the term $V(q=0)$, the sole effect of
this interaction in all that we have computed is to renormalize the
statistical parameter $\kappa$:
\begin{equation}
\label{ren}
\kappa\rightarrow\sqrt{\kappa^2
+{m\sqrt{2}\over{\bar\rho}\pi^{3\over 2}}V(0)}\,.
\end{equation}
For more general forms of $V(x-y)$ we can build a perturbation theory based
on the Luttinger model as the Fermi liquid is based on the ideal Fermi
gas \cite{Hald1}.

In summary, we described here an alternative description of
generalized statistics in 1D based on a gauge field theory that parallels
the Chern-Simons construction in 2D. It was shown that in the long wave
length  limit we have a correspondence to a gas with generalized exclusion
statistics, and the relation to 1D fermionic models was established.

The author is grateful to A.N.Vaidya, E.C. Marino and P. Gaete for many
stimulating conversations. This work was supported by the
CNPq (Brazilian Research Council).

\end{document}